# Electron-ion model of ball and bead lightning

Sergey G. Fedosin

Postal code 614088, Sviazeva str. 22-79, Perm, Perm Krai, Russia

E-mail: fedosin@hotmail.com

**Abstract:** Based on the electron-ion model, parameters of ball and bead lightning are calculated. The model allows us to estimate maximum size of ball lightning, its energy content, electric charge and magnetic field, to determine equilibrium conditions between positively charged ions located inside and outer shell containing rapidly moving electrons. An explanation is given to the fact that shells are observed inside ball lightning that differ from each other in color of glow. The model describes structure of not only ball lightning, but also bead lightning. The long-term stability of bead lightning is associated with the balance of neighboring beads under action of magnetic force of their attraction and electric force of repulsion, which exceed in magnitude the force of wind pressure.

## 1. Introduction

The relation between occurrence of normal linear lightnings and ball lightnings, as well as between occurrence of normal lightnings and bead lightnings is well-known. According to statistics, intensity of observations of ball lightnings (BL) repeats very well distribution of thunderstorm activity during a day and frequency of thunderstorms by months during a year almost in all countries. The same applies to bead lightnings, which usually occur as a result of transformation of normal lightnings into them [1-2].

In the model described in [3-4], it is concluded that BLs and bead lightnings appear in ionized air in channel of linear lightning, and at the moment of the lightning current termination they rapidly increase in size under action of an expansion wave directed radially with respect to the channel. After that, positive ions, located mainly in center of BL, and negative ions gradually annihilate, thereby reducing electric field inside BL. It is assumed that particles from the soil, captured by electric field, get inside BL and cause its subsequent glow, igniting in the very hot air located in lightning channel. The color of glow in this case depends on composition of the soil particles and amount of metals in these particles. The same particles may be responsible for sparks that appear near surface of BL. Despite the fact that fragmentation of linear lightning into separate segments is associated with the pinch effect, this model does not calculate the effect of magnetic field on the structure and configuration of currents and fields inside the emerging BL and segments of bead lightning. In addition, there is a difficulty in explaining the glow of those BLs that, according to observations, arise inside rain clouds where there are no soil particles.

In some models, it is assumed that there is a magnetic field inside BL, arising from the motion of charged particles of ionized plasma [5-6]. In [7], it is assumed that BL could be a plasmoid containing a force-free magnetic field, in which all electric currents in plasma are parallel to magnetic field lines. However, the magnetic field alone is not enough for a stable BL configuration; some additional forces are required [8-9].

There is quite a number of models that describe internal structure of BL. For example, in [10-11] a proton-electron model is developed, according to which the shell of BL contains high-energy protons, and electrons are located in the center of BL. Both protons and electrons are in rotational motion, creating a magnetic field of the BL. Due to the magnetic field, the Lorentz force arises, which, together with the electric force, maintains the stable rotation of charged particles inside the BL.

A characteristic feature of the model in [12] is that the ball lightning core contains electrons and almost fully ionized air, while electrons and ions are in oscillatory motion with respect to each other. Electrons oscillate mainly in the radial direction. The core is surrounded by a layer of heated air, which isolates it from the surrounding atmosphere. This model assumes that electrons in the ball lightning core can emit electromagnetic radiation in X-ray and gamma-ray ranges. The disadvantage of both models [10-11] and [12] is that there is no description of the processes that could lead to the proposed structure of BL, to the division of air molecules into the required composition of particles with the corresponding motion of these particles.

In [13-14], it is indicated that there is still no satisfactory model that describes all observed properties and mechanism of formation of bead lightning.

In this regard, in this paper we present general mechanism of formation and internal structure of both BL and bead lightning, based on the electron-ion model according to [15-19]. The charge configuration in this model is such that moving electrons and ions interact with each other by means of both electric and magnetic fields, ensuring long-term stability of BL.

Our purpose is to provide not only qualitative conclusions, but also quantitative calculations of main parameters of BL and bead lightning in their equilibrium state. The main attention will be paid to overall picture and consistency of results with physics of phenomena and with observation results. The dynamic aspects associated with phases of formation and disappearance of BL and bead lightning within the framework of electron-ion model, including charge recombination and ionizing effects, are described here briefly, since they require a separate article involving entire apparatus of ionized plasma physics.

Sections 2 and 3 will describe the mechanisms of occurrence and parameters of BL, and Section 4 is completely devoted to the description of occurrence of bead lightning with calculation of its physical parameters. In Section 5 we will compare the properties of BL in the electron-ion model with the results of BL observations, and in Section 6 we will compare the electron-ion model with the results of experiments and theoretical studies.

## 2. Features of electron-ion model

Within the framework of electron-ion model, natural BL is a direct consequence of linear lightning, when a thundercloud is discharged to the ground or to neighboring clouds. In model of a usual cumulonimbus cloud, described in [20] and confirmed by observations, upper part of the cloud is positively charged, and the lower part of the cloud is negatively charged. Apart from numerous discharges within clouds, linear lightnings are usually observed between the bottom of thunderclouds and the ground, and such lightning transfers negative electricity to the ground. As a thunderstorm cell develops, the lower part of the cloud can discharge, but the positive charge at the top of the cloud can remain. This positive charge can cause lightning, in which electrons move not from the cloud to the ground, but in the opposite direction – from the ground to the cloud.

In principle, both types of lightning, though they differ in direction of current in them, can lead to occurrence of BL. In case when electrons in linear lightning move towards a positively charged cloud, first, under the influence of strong electric field, the so-called step leaders are emitted from the ground towards the cloud, creating in the air conducting channels filled with electrons. Most of these channels will never connect with the cloud; rather, some of them will connect with each other and with the main lightning channel. Under additional conditions, such a multi-way configuration of branches with electrons in them turns out to be favorable for the occurrence of BL.

In another case, when electrons in lightning move from the negatively charged part of a cloud to the ground, the step leader moves towards the ground in jumps about 50 m long, at the velocity of about 1/6 of the speed of light and with stops of about 50 µs. When distance between the leader and the ground is about 100 m, positive charges start moving up from the ground towards the leader in the ionized air, creating conducting channels. As a result, the leader can branch out and connect with these channels, and the first of these connected channels will pass the greatest current through themselves and become noticeable due to bright glow.

In [15-17], [19], a mechanism is described in which BL occurs in the case when secondary branches of linear lightning form an almost closed circular configuration. If an electric current appears in main lightning channel, then the electrons present in a secondary branch begin to move into the main channel under influence of electric field. Due to circular configuration of the secondary branch, the motion of electrons creates a magnetic field that traps positive ions within the branch. When the current in the main channel suddenly stops, the electron current in the secondary branch can close on itself, forming outer shell of the emerging BL. Such a shell is held by the positive charge of ions inside BL. The described situation is illustrated in Figure 1.

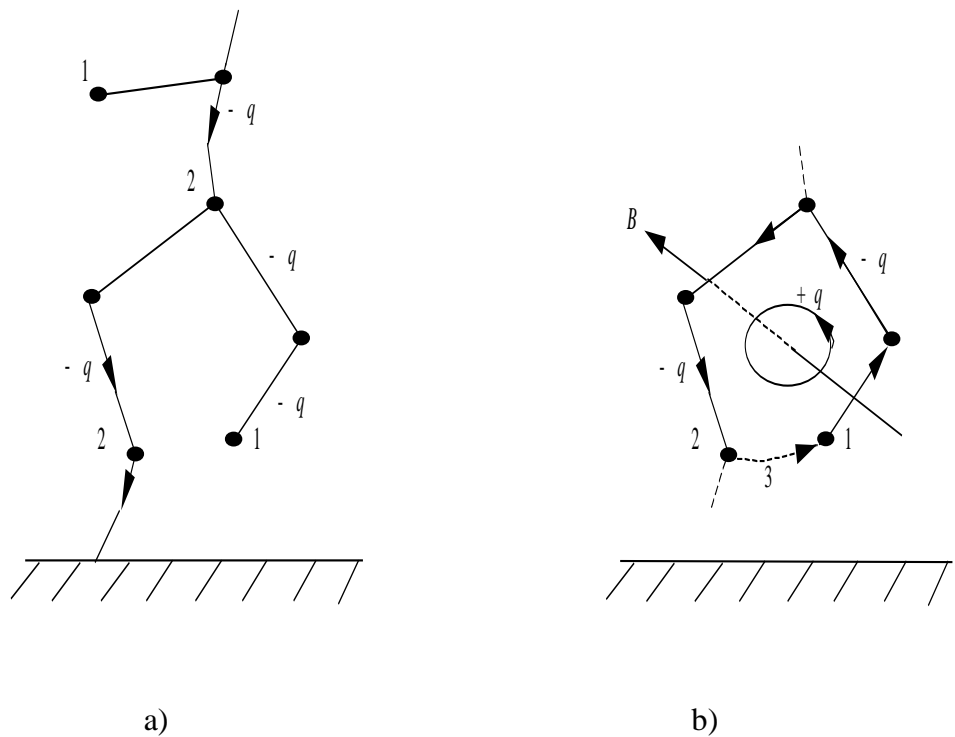

a)　　　　　　　　　　　　　　b)

**Fig.1. a)** 1 – secondary branches of lightning, 2 – main channel, in which electrons move (denoted by $-q$). **b)** The motion of electrons from the secondary branch (1) to the main channel (2) of lightning can be closed through region 3. *B* is magnetic field induction from the circular current of electrons. The ions with the charge $+q$ rotate relative to magnetic field lines in the same direction as electrons, but at a different angular velocity.

Another method of BL formation can be associated with the pinch effect, due to which the lightning current channel can be pinched by magnetic forces. While the electrons are moving in the lightning channel, they create a strong magnetic field, with magnetic force lines surrounding the channel and shaped like circles. Some electrons and ions near the lightning channel, if their velocities are directed along the magnetic force lines, move in circles around the channel. In the case when the motion of electrons inside the lightning channel stops, rotational motion of the external electrons around the channel can be maintained if there is a force that keeps the electrons from flying away. This force can be electrical force from positive ions inside the outer electron shell. The pinch effect is most clearly manifested when ordinary lightning is transformed into beaded lightning.

According to electron-ion model, BL has an axisymmetric configuration with a strong surface electron current in the thin outer shell. Positive ions can be under atmospheric pressure in the very hot air inside BL, left after a linear lightning stroke. Rapidly moving electrons in outer shell generate a magnetic field that keeps ions on orbits inside BL. Such BL will be stable as long as it maintains an equilibrium configuration of ion and electron currents isolated from each other.

In [21-22] we can find that the electron mean free path with respect to elastic collisions with neutrals in the air under standard temperature and pressure is within $1 – 1.7$ μm. With the electrons' velocity of the order of $10^6$ m/s this gives the time $10^{-12}$ s as an estimate of the time which it takes for one free electron to collide with an air molecule. The relative decrease in kinetic energy of an electron during one elastic collision depends on the ratio of masses of the electron and a molecule and is estimated as $10^{-4}$. This leads to the fact that after a series of collisions in about $10^{-8}$ s the electron loses its energy of motion and binds to some molecule, forming an ion. In this case, how can spherical electron currents exist for a long time, being in the outer shell of BL?

In order to answer this question, we will take into account that electrons at the moment of BL formation acquire significant energy through the electric and magnetic fields near the channel and branches of linear lightning. Due to this energy, electrons moving rapidly in the outer shell of BL continuously ionize the molecules of surrounding air, generating new fluxes of electrons that replace the losses of electrons. On the other hand, the total positive charge of BL acts on the positively charged air ions in such a way that it ejects them out of the outer shell with electrons, and then repels these ions from the surface of BL. Due to the high current density in the outer shell, electrons create their own dynamic pressure in this shell that prevents penetration of ions and air molecules into this shell. As a result, electrons, when moving in the outer shell, do not often collide with ions and do not lose their energy for a long time.

The high electric field strength of BL contributes to additional ionization of neutral air molecules. As for the interaction of the shell electrons with the air ions inside BL, this interaction occurs mainly through magnetic field, forcing the ions to rotate around the common axis of rotation. Fixation of ions by the magnetic field prevents them from interacting with the shell electrons. Although the radial electrostatic force of positive volume charge of BL also acts on the ions, the total pressure from this force on the shell is decreased due to the ions' connection with magnetic field, so that the force between volume charge inside BL and electrons in the outer shell can keep BL in equilibrium. Thus, the air ions inside BL rather weakly reduce the energy of motion of these electrons. All this effectively reduces the number of interactions between electrons in the shell and the air ions and molecules, as well as the loss of electron energy per unit time, which increases the lifetime of BL to observable values of up to tens of seconds.

The electrons in outer shell repel each other, but due to electrical attraction of positive ions inside BL, the electrons do not get scattered and are retained in this shell. In addition, the attraction of ions is the main part of centripetal force, which is necessary to hold electrons as they rotate in the outer shell [16-17]. Based on the approximately spherical shape of BL, the radius of rotation of outer electron cloud about common axis decreases as it moves from equator to poles. We can assume that the electrons in outer shell form electron liquid that uniformly fills the entire shell and is in continuous rotation.

This relatively stable configuration makes it possible to explain the observed lifetime of BL, which is much longer than lifetime of homogeneous ion-electron plasma under atmospheric pressure. The electron shell effectively isolates the air heated to high temperature inside BL, slowing down the energy transfer to environment.

## 3. Physical properties of ball lightning

Electric currents and magnetic field inside the BL are shown in Figure 2.

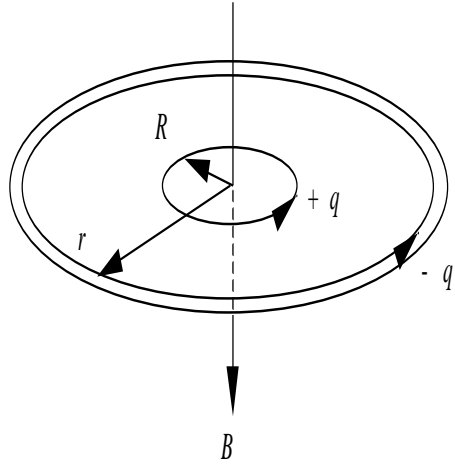

**Fig. 2.** Equatorial section of ball lightning, highlighting rings on spheroidal current shells. $R$ is the radius of ions' rotation in equilibrium shell around magnetic field lines with induction $B$, $r$ is the radius of outer electron shell.

Inside BL we can find shells where thermal velocity of certain ions is equalized with the velocity of rotation of these ions in magnetic field [17]. At this point, we can equate the centripetal force acting on rotating ions and the Lorentz magnetic force acting on the ions in the equatorial plane perpendicular to magnetic field:

$$\frac{MV^2}{R} = qVB, \qquad (1)$$

where $M, V, R$ are the masses, velocities, and radii of rotation of ions,

$q$ is elementary electrical charge, $B$ is magnetic field induction.

In the general case, instead of (1), we should write a system of vector equations of motion of neutral and charged particles (air molecules, ions and electrons) in electromagnetic field, taking into account vector pressure field and vector acceleration field similarly to [23-24]. Another approach involves the use of equations of hydrodynamics and magnetohydrodynamics [5], [21] with corrections for various effects of electromagnetic field's action on the air molecules. Instead, we isolate the leading terms of the equations to obtain estimates of physical quantities in the first approximation.

In order to simplify calculations, it is assumed in (1) that most of ions are ionized once, and therefore their charge is equal to elementary electric charge. Indeed, according to [25], the observed BLs contain singly ionized oxygen and nitrogen atoms, as well as doubly ionized nitrogen.

On the other hand, charged particles can freely move along magnetic field lines, passing along the axis of BL from one pole to the other. Consequently, there could also be such shells inside BL, where ordered mutually perpendicular ion fluxes take place, which is accompanied by strong friction in gas and corresponding energy release in the form of radiation. Since the air consists of several gases, then positive ions of various masses and various degrees of ionization appear at high temperatures. These ions would have not coincident thermal velocities and radii of rotation in magnetic field. As a result, nested luminous shells of different colors and brightness can appear in BL, depending on composition of ions and on the rate of transition of kinetic energy into radiation during collisions of ions in one shell or another.

Let us present our calculations of parameters of BL with the total radius $r = 7$ cm and the radius of isolated shell $R = 4$ cm, similar to that described in [26-27]. For the sake of simplicity, we will assume that charges and currents are mainly concentrated near the equatorial plane or have cylindrical symmetry, and ions are singly charged. According to [17], the electron current $I$ in outer shell and magnetic field $B$ of this current acting on the ions can be expressed in terms of number of electrons $N_e$ and their velocity $v$ at the known temperature $T$:

$$I = \frac{qN_e v}{2\pi r}. \tag{2}$$

$$B \approx \frac{\mu\mu_0 I}{2r}. \tag{3}$$

$$\frac{mv^2}{r} = \frac{qQ}{4\pi\varepsilon\varepsilon_0 r^2}. \tag{4}$$

In (2) $qN_e$ is total charge of free electrons in outer shell of BL, the quantity $\frac{2\pi r}{v}$ defines period of the electrons' rotation, so that the current $I$, in accordance with its definition, is obtained as the ratio of entire moving charge to time equal to period of the electrons' rotation. Relation (3) describes magnetic field $B$ on the axis of ring current $I$, with the radius of the ring equal to $r$, $\mu$ is the magnetic permeability of medium inside BL, $\mu_0$ denotes the vacuum permeability.

In (4) the condition for the equilibrium of an electron during its rotation in the electron shell of the BL is presented, while $Q = q(N_i - N_e)$ is a total charge of the BL, $N_i$ is the number of positively charged ions inside the BL, $m$ is the mass of the electron $\varepsilon$ is a relative dielectric permittivity of the medium inside the BL, $\varepsilon_0$ is the electrical constant.

In order to use the condition of equilibrium of forces for ions (1), it is necessary to estimate the average mass $M$ of a typical air ion and velocity $V$ of its motion, which is equal in the order of magnitude to velocity of thermal motion. The relation between kinetic and thermal energies for an ion has the form

$$\frac{MV^2}{2} = \frac{3\kappa T}{2}, \tag{5}$$

where $\kappa$ is the Boltzmann constant.

An approximate relation for magnetic field in the center of BL follows from (1) and (5):

$$B \approx \frac{\sqrt{3\kappa T M}}{qR}. \tag{6}$$

With the temperature of air inside BL of the order of $T = 1.4 \times 10^4$ K and with the average ion mass $M = 4.7 \times 10^{-26}$ kg, like that of a nitrogen molecule, the velocity of ions in (5) will be equal to $V = 3.5 \times 10^3$ m/s. Taking into account that in this BL $r = 7$ cm, $R = 4$ cm, from (6) and (3) at $\mu \approx 1$ we find the magnetic induction inside the BL under consideration, which is equal to $B \approx 0.026$ T, and total current of electron shell $I \approx 2900$ A.

Based on the approach considered in [17], other parameters of BL were found depending on its radius. Thus, for BL with a radius $r = 7$ cm, an estimate of number $N_e$ of electrons in the outer shell and number $N_i$ of ions inside BL gives $N_e \approx N_i \approx 9.2 \times 10^{13}$; total electric charge of BL $Q = q(N_i - N_e) = 2.9 \times 10^{-7}$ C; velocity of electrons' motion in the outer shell $v = 8.7 \times 10^7$ m/s; kinetic energy of electrons $E_k = \dfrac{N_e m v^2}{2} = 0.3$ J; magnetic field energy 0.4 J; electrostatic energy in the volume of BL is 2.8 J; internal energy of ionized air inside BL is 500 J; electric field strength and electric potential near the surface of BL are 530 kV/m and 37 kV, respectively.

The pressure from electrical repulsion of ions inside BL is equal to $P_i = 2.8 \times 10^4$ Pa, and magnetic pressure acting on the ions is equal to

$$P_{mi} = \frac{B^2}{2\mu\mu_0} = 2.7 \times 10^2 \text{ Pa.} \tag{7}$$

The magnetic pressure $P_{mi}$ and the pressure $P_i$ from the electrical repulsion of ions inside BL do not exceed the atmospheric pressure, so the outer shell with electrons, which is under the external atmospheric pressure, can retain the ionized and charged air inside BL.

Let us consider rotation of electrons in the outer shell near equator of BL. The electrons here are under action of centripetal force in the form of electrical attraction from the positive ions inside BL. If we take into account only this force, then the virial theorem is satisfied in its simplest form, when the sum of doubled kinetic energy of electrons and their potential energy in electric field is equal to zero. However, the equation of electrons' motion in accordance with the plasma equilibrium condition in magnetohydrodynamics includes two additional small forces, one of which is proportional to pressure gradient in the electron shell, and the other is Lorentz magnetic force [21]. The Lorentz force can be transformed into s term containing gradient and also into a term that can be neglected in the first approximation. This is how the relation $P_{me} + P_{\perp} = const$ appears, where $P_{me}$ is magnetic pressure in the shell with electrons, $P_{\perp}$ is electron pressure across the magnetic lines. The electrons moving in outer shell at velocity $v$ create a pressure equal in the order of magnitude to

$$P_e = n_e k T_e = \frac{N_e k T_e}{4\pi r^2 d}, \tag{8}$$

where $T_e$ is kinetic temperature of electrons, $W_e = \frac{3\kappa T_e}{2} = \frac{mv^2}{2} = 3.4 \times 10^{-15}$ J is kinetic energy of one electron, $m$ is electron mass, while the energy $W_e$ is 22 keV in energy units. If we assume that $P_e = P_\perp = P_{me} = P_{mi}$, then in view of (7-8) we obtain the thickness of BL shell which equals $d = 1.3$ cm.

In the case of BL formation in linear lightning channel, the electron velocity distribution may be such that electrons will have the velocity $v$ not only in outer shell, but also in central part of BL. In this case, such electrons will rotate in magnetic field $B$ and will be held at the center of BL. From the equality of centripetal force and Lorentz force we obtain the following relations:

$$\frac{mv^2}{r_e} = qvB, \qquad r_e = \frac{mv}{qB} = 1.9 \text{ cm}. \tag{9}$$

The radius $r_e$ in (9) limits central part of BL, filled with fast moving electrons. These electrons can provide that violet glow in the center of BL that was observed in [26-27]. Comparing volume of central part of BL with the radius $r_e$ and volume of outer shell with the thickness $d$, we can conclude that energy density of electrons near the center of BL can be by an order of magnitude higher than energy density of electrons in the outer shell of BL. This allows us to explain the fact that when BL passes through glass, holes with a diameter ranging from fractions of a millimeter to the size of BL itself are usually observed [2], [28]. Sometimes holes can be melted, sometimes not; in the latter case, it can be assumed that the glass first gets very hot and then breaks off under pressure of BL moving under influence of strong electric fields.

Table 1 shows calculated parameters for BL of different sizes. Designations in the table: $r$ is radius of BL, $I$ is total electron current in BL shell, $B$ is magnetic field induction in the center of BL, $W$ is total energy taking into account the energies of magnetic and electric fields, the energy of electrons' motion and the plasma energy inside BL.

**Table 1.**

**Parameters of ball lightnings of different sizes**

| $r$, cm | $I$, A | $B$, T | $Q$, C | $W$, J |
|---|---|---|---|---|
| 1 | 20 | $1.3 \times 10^{-3}$ | $4 \times 10^{-12}$ | 2.2 |
| 7 | $2.9 \times 10^{3}$ | $2.6 \times 10^{-2}$ | $2.9 \times 10^{-7}$ | 503 |
| 17 | $1.4 \times 10^{5}$ | 0.5 | $9.6 \times 10^{-6}$ | 10600 |

The radius of the smallest BLs is about 1 cm, while the energy of electrons' motion and magnetic energy can be considered approximately the same. At a radius of 7 cm, the energy of BL is equal to 503 J, almost all of it is contained in thermal energy of moving ions and in potential energy of ionized particles.

The main energy of powerful BL with its radius of 17 cm is contained in energy of ionized particles and in energy of electromagnetic field. In this case, the total energy of 10.6 kJ falls within the range of upper values of observed energy content of BL, determined through interaction of BL with surrounding objects (melting of metal parts, heating and evaporation of water, etc.).

In the model under consideration, BL in general must be positively charged so that the ions inside BL could retain the electron shell for a long time. However, the total charge of BL cannot exceed such a value, at which electric field strength on its surface exceeds $E_0 = 3000$ kV/m, in order to avoid the breakdown of atmospheric air. Due to high electric field strength and rapid electrons' rotation in BL shell, the observed effects of hissing, crackling and sparking occur. The observed radius of BL can be greater than 17 cm in Table 1 due to glow of electrified air. In addition, there may be burning soil particles in the air, trapped in the BL's electric field.

As can be seen from Table 1, the maximum electron current in BL is quite close to current in the channel of average linear lightning.

### 4. Bead lightning

When constructing the bead lightning model, we will rely on the data from [29]. Using frame-by-frame reconstruction of a video film, which recorded lightning striking spire of Ostankino TV tower in Moscow on July 24, 2001, formation of bead lightning was discovered at the site of decay of normal lightning. The average distance between the neighboring beads (or "pearls") of the bead lightning was $z = 1$ m, the average bead length was $L = 0.65$ m, and the bead width when viewed from the side was $D = 0.41$ m. Since the width of the primary linear lightning channel was 0.25 m, the bead width $D$ exceeds the width of the primary lightning channel.

We can assume that each bead has the shape of a cylinder with the length $L$ and diameter $D$, and the area of maximum lateral section of such a cylinder is $S = LD = 0.27$ m². Since wind speed at the height of the spire of 533 m on that day equaled $V = 43$ m/s, and dynamic air pressure could be expressed by the formula $P = \dfrac{\rho V^2}{2}$, where $\rho = 1.3$ kg/m³ was the air density, then the pressure force from the wind on each bead had the following value:

$$F = PS = \frac{\rho V^2 S}{2} = 324 \text{ N}. \tag{10}$$

The paradoxicality of the situation with that bead lightning was that despite the hurricane wind, it almost did not move during its lifetime of 0.625 seconds. The video recording showed that the lightning beads remained in their places. The authors of [29] suggested that the lightning beads could possibly be held by electric forces; for this case, taking into account (10), they calculated the corresponding electric field strength $E$:

$$E = \frac{Q}{4\pi\varepsilon_0 z^2}, \qquad F = QE, \qquad E = \sqrt{\frac{F}{4\pi\varepsilon_0 z^2}} \approx 1.7 \times 10^6 \text{ V/m}, \tag{11}$$

where $\varepsilon_0$ is vacuum permittivity, $Q$ is charge of a bead.

The electric field strength $E$ in (11) does not exceed the air breakdown strength $E_0 = 3 \times 10^6$ V/m.

However, the point is that free charged objects cannot be fixed in their places by the electrical forces alone – they will be unstable, and either rush towards each other or scatter in different directions, depending on the signs of charges.

Let us apply to analysis of bead lightning the data on structure of ball lightning (BL) presented in the previous section. According to the electron-ion model, BL has a positive charge, as well as magnetic field that arises mainly from electrons rapidly rotating in the outer shell of BL. Let us assume that the structure of individual beads of bead lightning is the same as that of BL, then the neighboring beads are repelled from each other by electric forces, and at the same time are attracted by magnetic forces. As a result, the electromagnetic forces are balanced, and the beads remain in their places until the complete loss of energy in the lightning.

Figure 3 shows two neighboring beads interacting with each other. The inherent stability of each bead is determined by the fact that internal volumetric positive charge holds the rotating electrons in the bead shell by electric forces. At the same time, magnetic field arising from the motion of electrons fixes positive charges inside the bead as they rotate in this magnetic field.

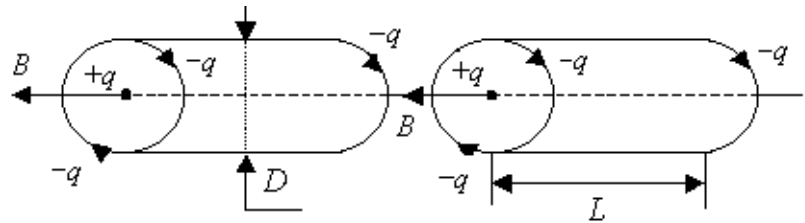

**Fig.3.** The structure of beads in bead lightning in the form of cylinders. Electrons (denoted by $-q$) rotate on surface of beads and create magnetic field with induction $B$, inside the beads' volume there is hot ionized air with positive charges $+q$. $D$ is the beads' diameter and $L$ is the beads' length.

Let us now consider stability of the beads relative to each other. If each bead, just like BL, has the total charge $Q$, then the electric repulsion force of equally charged beads equals:

$$F_e = \frac{Q^2}{4\pi\varepsilon_0 z^2}, \qquad (12)$$

where $z$ is distance between the neighboring beads.

The magnetic force of attraction between the beads is approximately calculated as the force of interaction of two identical solenoids located on the same axis at the distance $z$ from each other:

$$F_m = \frac{\mu_0}{4\pi} \cdot \frac{\left[\dfrac{I}{L} \cdot \dfrac{\pi D^2}{4}\right]^2}{z}. \tag{13}$$

In (13) $I$ is electron current through the bead's cylindrical surface, the ratio $\dfrac{I}{L}$ is current per bead's unit length, the value $\dfrac{\pi D^2}{4}$ is cross-sectional area of the bead's cylinder, $\mu_0$ is vacuum permeability.

If we do not take into account the action of remaining beads, the balance of two neighboring beads is achieved when the forces $F_e$ (12) and $F_m$ (13) are equal, which, in view of the well-known identity $c^2 \varepsilon_0 \mu_0 = 1$, gives the following:

$$\frac{Q^2}{4\pi \varepsilon_0 z^2} = \frac{\mu_0}{4\pi} \cdot \frac{\left[\dfrac{I}{L} \cdot \dfrac{\pi D^2}{4}\right]^2}{z}, \qquad Q = \frac{I}{cL} \cdot \frac{\pi D^2}{4}, \tag{14}$$

where $c$ is the speed of light.

The magnitude of electron current $I$ can be related to the number of electrons $N_e$ in the bead's cylindrical shell by corresponding formula, so that with the electrons' velocity tending to the speed of light, for the current, similarly to (2), we find:

$$I = \frac{q N_e c}{\pi D}, \tag{15}$$

where $q$ is elementary electric charge.

The fact that the maximum velocity of electrons' motion in (15) can be of the order of the speed of light is due to extremely high current intensity in the lightning that struck the spire of Ostankino TV tower in Moscow on July 24, 2001. At the temperature in lightning of the order of $T = 25000$ K, according to [30], the thermal velocity of electrons, calculated by formula

(4) taking into account electron mass $m$ instead of ion mass $M$, gives the value $V_e = \sqrt{\dfrac{3\kappa T}{m}} \approx 10^6$ m/s. On the other hand, electrical forces can accelerate electrons more efficiently than thermal heating of gas. This is apparent from the fact that step leader during lightning formation can move at the velocity of the order of 1/6 of the speed of light [20]. Thus, formula (15) gives an estimate of maximum possible electron current $I$. More precise current values could be determined if we take into account relativistic corrections and conditions of current flow in ionized plasma and in outer shell of bead lightning.

Substitution of the current (15) into relation (14) gives:

$$Q = \frac{qN_e D}{4L} \qquad (16)$$

On the other hand, the total charge of one bead $Q$ is expressed as difference between all positive and negative charges, then, in view of (16), we have:

$$Q = q(N_i - N_e) = \frac{qN_e D}{4L}, \qquad N_i - N_e = \frac{N_e D}{4L} = 0.16 N_e, \qquad N_i = 1.16 N_e, \qquad (17)$$

so that the number of positive ions $N_i$ inside the bead is 1.16 times greater than the number of electrons $N_e$ in the bead shell.

Now we will equate the wind force $F$ from (10), acting on the bead, to the electric force $F_e$ from (12), which will allow us to find the charge of the bead:

$$F = F_e = \frac{Q^2}{4\pi\varepsilon_0 z^2}, \qquad Q = z\sqrt{4\pi\varepsilon_0 F} = 1.9\times 10^{-4} \text{ C}. \qquad (18)$$

Substituting the charge (18) into (16), we determine the number of electrons in the bead shell:

$$N_e = \frac{4LQ}{qD} = 7.5\times 10^{15},$$

as well as the number of uncompensated positive ions inside the bead, according to expression (17): $N_i = 1.16 \, N_e = 8.7 \times 10^{15}$

Knowing the quantity $N_e$, we can estimate the electron current in the bead shell from (15):

$$I = \frac{qN_e c}{\pi D} = 2.8 \times 10^5 \text{ A}. \qquad (19)$$

The magnetic field induction reaches its maximum in the center of the bead and based on the formula for solenoid is equal to:

$$B_m = \frac{\mu_0 I}{\sqrt{D^2 + L^2}} = 0.46 \text{ T}. \qquad (20)$$

Comparison of obtained values of internal induction of magnetic field and electron current in the bead shell with the corresponding values for powerful BL from Table 1 shows that they are close in magnitude. Based on the magnitude of magnetic field, we can conclude that magnetic pressure in the center of the bead, determined by the formula $P_m = \frac{B_m^2}{2\mu_0}$, does not exceed atmospheric pressure in the air outside the bead volume. This means that due to the magnetic pressure, the volume of the bead has an ability to change within certain limits, adjusting to forces acting in the bead.

Apparently, formation of bead lightning from normal lightning occurs due to the pinch effect, in which the conductive channel of lightning in some place is pinched by the magnetic forces arising from the flowing current. Indeed, according to Ampère law, the parallel electron currents in lightning channel are attracted to each other. Due to the current in the lightning, the magnetic field appears, the force lines of which have the form of circles around the current channel. As a result, the lightning current channel is surrounded by the shell of electrons rotating along the magnetic lines and shielding the channel from external environment. The greater is the current in lightning and the greater is magnetic field, the smaller is distance from magnetic lines, at which the shell's electrons can move off during their motion. On the average, this distance does not exceed the radius of Larmor rotation due to Lorentz force.

Under condition of compensation of electric forces due to simultaneous presence of positive and negative charges and with the sufficient current, magnetic pressure can exceed the gas-dynamic pressure and lead to restructuring of the lightning structure — translational motion of charges along the entire lightning channel becomes difficult and rotational motion becomes significant in individual formations, that is in the beads of lightning. The condition for occurrence of pinch effect is described in this case by the Bennett relation [31], which takes into account equality of densities of current's magnetic energy and thermal energy of electrons and gas ions:

$$\frac{\mu_0}{2}\left(\frac{I}{2\pi r}\right)^2 = n\kappa(T_e + T_i). \tag{21}$$

If we substitute in (21) typical values of parameters for linear lightning according to [30], in particular, radius of lightning channel $r = 0{,}1$ m, concentration of electrons $n = 4\times 10^{24}$ m$^{-3}$, electron and ion temperatures $T_e \approx T_i \approx 25000$ K, and take into account values of magnetic constant $\mu_0$ and Boltzmann constant $\kappa$, then we can estimate the current magnitude: $I = 1{.}3\times 10^6$ A. As could be expected, this current is quite strong and not less than the current (19) in the bead shell. Current strength in lightning that struck the spire of the Ostankino television tower in Moscow on July 24, 2001 could really be so significant, since the length of the lightning and accordingly electrical resistance of lightning channel decrease due to height of the television tower.

Defragmentation of linear lightning into separate beads begins with a decrease in current strength in lightning channel, due to which external magnetic field decreases, and electron shell moves away from the channel. Since motion of electrons in lightning shell creates a longitudinal magnetic field in the channel, this field, magnetic pressure from which counteracts the occurrence of pinch effect, also decreases. With expansion of electron shell of lightning, the volume of lightning channel in some places increases to the size of beads, while the channel takes form of separate beads with gaps between them, mainly as a result of electric repulsion of the positively charged contents of neighboring beads.

As in the case of ball lightning, the beads' glow can be associated with the friction of mutually perpendicular ion fluxes, one of which is directed along magnetic field lines inside the beads, and the other is associated with rotation of ions around these magnetic lines.

According to [32], the fraction of free electrons in the air is close to 0.7 at the electric field strength of the order of 700 kV/m. Inside and outside the beads, the field strength can be even greater than 1700 kV/m according to (11). Such fields lead to appearance of a large fraction of free electrons, which makes it possible to substantiate the number $N_i$ of positive ions inside the bead and the number $N_e$ of electrons in the bead shell in (17) within the framework of electron-ion model. On the other hand, electric field near the beads has the same order of magnitude as the critical electric field $E_0 = 3000$ kV/m. This leads to a rapid loss of energy by bead lightning due to discharge into the surrounding air and to a decrease in the lifetime of bead lightning by an order of magnitude as compared to the lifetime of ball lightning.

### 5. Compliance of electron-ion model with observations

As indicated in Section 2, BL can occur in curved secondary branches of linear lightning when sufficiently large circular electron currents appear in these branches. Such currents generate a dipole magnetic field that retains positive ions inside the resulting BL. The result is a stable configuration, shown in Figure 2. In this configuration, the electron shell contains fast-moving electrons, which are subject to an electric force from the positively charged ions inside the BL.

As an example, [33] describes a case when one of the branches of linear lightning, passing almost horizontally near a power transmission line, started twisting in an arc and BL immediately appeared at this place. In a number of other cases, BL occurred when two different lightnings or their branches crossed each other, as well as at the places of lightning branching.

Another option for BL formation is associated with the pinch effect, the condition of which is presented in (21). This mechanism is supported by the fact that sometimes not one but two BL are found at once, connected to each other by a chain of luminous beads, similarly to connection of beads in bead lightning [30]. Occasionally, BL with a shape of an elongated cylinder are observed [2]. In [33], the case was observed in which, after a linear lightning strike, the lightning channel began to fade out, consistently losing its glow, starting from the clouds towards the ground. When only a small part of the luminous channel remained near the ground, BL appeared at this place, which seemed to be rapidly rotating.

As an illustration, we can cite a case with two researchers according to [33], in which during a thunderstorm in the mountains at an altitude of 2700 m, such a strong electric field emerged that their hair became electrified and stood on end. When a lightning strike occurred at the distance of 25 m from the observers, at the point where the lightning hit the ground, a ball began to twist and grow, which eventually turned into BL. Then BL jumped up 35 cm above the ground and started floating off to a side. When BL clung to stones, sparks were flying and a sound was heard, similar to metal ringing and knocking of a baby rattle at the same time. In other similar cases, the BLs remain close to the ground or move at a low altitude.

Based on such observations, it is concluded that BL can be formed in the linear lightning discharge channel at the stage of its decay through an intermediate state such as beaded lightning. In addition, BL can also occur in secondary branches of linear lightning, in which there can be a significant electric current even in the absence of noticeable glow. Observations from airplanes show that many BLs occur inside cumulonimbus clouds, which is associated with abundance of branches of each linear lightning, and with a large number of electrical discharges inside and between clouds.

Thus, BL can appear even away from the brightly sparkling main channel of linear lightning at any height from the ground. In some cases, the distance between emerging BL and main channel of linear lightning can reach hundreds of meters [34], which can be explained by the branching of main lightning channel into several less bright channels. Then BL emerges near one of the secondary channels, and really, according to observations, many BLs emerge near relatively weak lightning with current of the order of tens of kA.

In [35], the cases are described of even greater distances between BLs occurring during a thunderstorm and those points on the ground where strikes of normal linear lightnings were recorded. In order to explain this situation, it should be taken into account that large neighboring cumulonimbus clouds can have opposite signs of their charges. Thus, one cloud can be charged negatively in the lower part, and positively in the upper part, and another cloud as a whole can be charged positively due to a significant discharge of its negative charge, which occurred earlier due to many lightning strikes. Then the first cloud will produce standard thunderstorm activity, including intra-cloud lightnings, rain showers and normal negative cloud-ground lightnings with electrons moving from the cloud to the ground. In contrast, the second cloud, moving several kilometers from the first, will likely generate positive cloud-ground lightnings with electrons moving from the ground to the cloud, and BL accompanied by light rain. In such a case, the probability of BL occurrence during positive lightnings significantly exceeds the probability of its occurrence during negative lightnings.

In [30], the cases are described when spherical objects appeared at the end of linear lightning descending from a cloud, but not yet reaching the ground. Such linear lightnings can be positive lightnings, in which electrons move from the ground to a positively charged cloud. When the lightning's conductive channel reaches the cloud, the electrons near the cloud are the first to start moving in the channel. Their current causes the lightning channel to glow, which spreads down to the ground. In this case, electrons from the ground flow into the channel from all sides like streams, forming luminous spherical object at the end of lightning moving towards the ground. This object is fueled by the energy of linear lightning and it can be so powerful that it can completely split a huge tree [33]. There is a possibility that if the cloud manages to discharge before the luminous channel reaches the ground, the current in lightning will stop, and glowing object at the end of the lightning will turn into free BL. We can assume that calculations of the energy of very powerful BLs for 17 observation cases in [36] refer specifically to such objects.

According to [37-38], out of 67 direct observations of BL occurrence, 31 events occurred in close proximity to linear lightning channel, 29 events were associated with "blowing-out" of BL from metal conductors such as electrical sockets, radio receivers, antennas and telephones, and in the remaining 7 cases BL appeared "from nothing". Quite often, BL occurs in an enclosed space, while almost simultaneously or within a short period of time about a minute earlier or later, a lightning strike occurs somewhere very closely [2], [30], [39]. All this indicates that in the room and close to it, under the action of strong electric field of cumulonimbus cloud, electricity-conducting branches are formed in ionized air, which are ready to transform into future lightning's channel. Most commonly this should happen in those premises that don't have high lightning rods. Metal conductors indoors conduct electricity better than ionized air, therefore most often BL occurs near these conductors at the moment of electrical breakdown of air and subsequent impulse of strong electric current. Indeed, cases have been recorded when during opening and closing of electrical circuits of battery units and powerful generators strong currents appeared and BLs occurred [2], [30].

In [26-27], the researcher Dmitriev provided a detailed description of a BL observed by him with the diameter of about 14 cm in horizontal direction and with the vertical dimension of up to 16 cm. According to this description, the inner luminous shell (with brightness similar to that of plasma heated in a plasmatron to 16000 C) was elongated in the vertical direction, and all the shells were clearly visible only in the horizontal direction. Separate bright convective jets were indicative of rotation of particles in the shells at a differential velocity. The central part of the lightning was surrounded by an area 1 – cm thick with dense violet glow, very similar to the air glow at a pressure of 0.1 mm Hg, bombarded by electrons with an energy of several tens of electronvolts. The next outer shell about 2 cm thick was also inhomogeneous, resembling in color a quiet electric discharge at atmospheric pressure or a peripheral glow of an electron beam with an energy of several tens of kiloelectronvolts, which comes into the air from a vacuum tube at ordinary pressure. The light blue glow of this part of the lightning quickly decreased with increasing of distance from the central ball, gradually fading away. The presence of high electric field strength near BL and high-energy particles was confirmed by observation of their hissing, crackling and sparking as during an electric discharge. In addition, the air samples taken after the passage of BL showed an increased content of ozone and nitrogen oxides (approximately 50-100 times higher than the norm). The required concentration ratio of ozone and nitrogen oxides can be achieved with an electric discharge in the air with a strength of up to 400 kilovolts per 1 meter, and an estimate of the required electric energy in such an equivalent discharge for the entire lifetime of BL gives a value of 530 J.

In the electron-ion model, it becomes clear why in Dmitriev's ball lightning all the shells were clearly visible only in the horizontal direction. Apparently, in this lightning, the axis of electrons' rotation in outer shell and internal magnetic field were directed vertically, so that there was a flux of ions moving along the magnetic field lines. This flux, when interacting with ions rotating in magnetic field, formed the shells observed in horizontal direction.

The parameters presented in Section 3 describe BL, which was observed in [26-27], quite well. The presence of strong electric current in the shell of BL allows us to explain observation described in [33]. During a thunderstorm, BL of about 6 mm in diameter was formed on contacts of transformer of overhead telephone line with a crackling and hissing sound. BL approached one of the two telephone line switchboards located in the room and discharged to its relay. At the same time, all 160 calling lamps of both switchboards flashed brightly, despite the fact that the operating voltage of the lamps was 24 volts. Apparently, when BL discharged to the contacts, it could become a source of electromotive force in the circuit connecting the lamps. In another example, BL discharged between the TV-set and the heating battery, and the kinescope of the TV-set suddenly lit up, although the TV-set was turned off.

In [14], [40-41] almost half a thousand cases of BL observations were analyzed, according to which the diameter of the most frequently occurring BL is in the range of 25-30 cm. At the same time, in almost 50 % of cases the BL diameter was less than 25 cm. Statistical dependencies showing the distribution of several thousand observations depending on the diameter of BL can be found in [42-44].

According to [45], the correlation parameter for observations of BL diameter in large databases reaches the value of 0.99. Description of observations using the log-normal distribution gives the standard deviation from the average value equal to 16 cm for the diameter and 11 seconds for the lifetime of the BL. In this case, the average value of the BL diameter for 4219 observations is equal to $28 \pm 4$ cm. The most probable energy value of the observed BLs lies in the range from 4.4 to 11 kJ. The above distributions are in good agreement with the fact that in the electron-ion model the maximum diameter of the BL in Table 1 does not exceed 34 cm and may be slightly larger taking into account the halo. The energy of such BL is equal to 10.6 kJ.

**6. Correspondence of electron-ion model with theoretical calculations and experiments**

The occurrence of BL in the place where the current lines have a bend is confirmed by the calculation in [46] based on the interaction of the current in the linear lightning branch with the magnetic field of this current. Under the influence of electromagnetic forces, a curved branch of ionized air with current will tend to turn into a ring, which then turns into a spherical shell of the emerging BL.

The initial stage of BL formation as a result of the pinch effect, using the equations of magnetohydrodynamics [13], [21], is described in [5], [47]/

The breakdown of air at the moment of opening of circuit with a strong current, or breakdown of air near electrical circuits' contacts under influence of high voltage gives a break in current line similar to a break in branches of linear lightning channel. Near such breaks, due to sharp turn of the current line, an additional magnetic field appears, which is one of the conditions for BL occurrence. This dipole-type magnetic field is directed perpendicular to the plane, in which the current lines or the lightning channel lines are located, has its own symmetry axis and thus differs from the magnetic field of direct current.

Due to such a magnetic field, positive ions, which appear at the breakdown site as a result of ionization by current or voltage, start rotating around the field symmetry axis, while an electron shell quickly grows around them. Electrons are moving along the current lines in crossed fields, one of which is the electric field from ions, and the other is the dipole magnetic field. At the moment when the main current stops, motion of electrons around the region with positive ions continues due to confining electric force, and BL is formed with a closed spherical electron current in its shell.

One example of mathematical description of BL formation in a high-voltage discharge near two contacts, one of which is cathode and the other is anode of an electrical circuit, is article [48]. At the initial moment of discharge, a streamer appears, which can branch both near the cathode and near the anode. Next, the streamer leads to formation of a plasmoid, which can turn into BL. The calculations use a streamer model modified with account of diffusion of electric field and ions. For the purpose of 3D modeling of the process, a typical unit of length is selected (as the mean free path of electrons in nitrogen under normal conditions), as well as typical units of time, electric field, electrons' density and velocity. A dispersion relation is derived that relates the wave numbers, electron diffusion coefficient, electric field strength and streamer velocity. This makes it possible to plot a graph of electron density distribution in the streamer and in the plasmoid, the potential and the electric field distributions, and other quantities depending on time. The typical time of formation of a 20 cm plasmoid ranges from 150 to 500 ms, while the minimum required electric field strength reaches the value of 35-40 kV/m [49]. Electric fields would have the same order of magnitude in BLs if they occur near electrical sockets, radio receivers, antennas and telephones located in closed rooms and exposed to high voltage from lightning.

In [50], it was experimentally shown that the luminous low-temperature plasmoids created during a discharge in a laboratory have a thin shell containing negative charges and preventing mixing of the plasmoid contents with atmospheric air. When plasmoids come into contact with conductors, the latter, as a rule, melt, and interaction with dielectrics is much weaker. Plasmoids also react to a laser beam, have an uncompensated electric charge, while the air temperature inside such artificially created plasmoids is about 200°C. However, the color temperature of plasmoids can reach 2000°C, which indicates the non-thermal nature of radiation.

It was found in [51] that there is a sharp mass density gradient in the outer shell of plasmoids. According to observations in [25], natural BL turned out to be sensitive to high-voltage (35 kV) transmission lines located close to it, since the frequency of changes in the glow intensity of BL coincided with the frequency of 50 Hz of electric current in power transmission lines. It is also well known that appearance of BL leads to appearance of noise in a working radio receiver.

All these facts can be explained as follows. In the electron-ion model, positive ions inside BL are weakly attracted by electrons from the outer shell, since electric field of this shell inside the sphere is close to zero because of balancing of all the electric forces due to the sphere's symmetry. Indeed, a charge uniformly distributed over the surface of a hollow sphere does not create an electric field inside the sphere, as follows from the Gauss theorem. Therefore, ions can be fairly freely distributed in the volume inside BL, reaching the state of equilibrium rotation in the internal magnetic field, while recombination of ions and electrons in BL is significantly slowed down due to spatial separation of charges. Electrons moving rapidly in the shell of BL create a strong current capable of melting conductors, burning wooden objects, heating and evaporating water. Interaction of electrons with the air surrounding BL leads to its ionization and is accompanied by radio radiation, which is perceived by radio receivers as noise. Being an object with electrically connected spatially separated positive and negative charges, BL reacts to the laser beam, since the photons act on electrons in the outer shell. The presence of charge in BL ensures interaction of this charge with the alternating electric field near power transmission lines, which is expressed in a change in brightness of BL glow with the frequency of current in these lines.

According to [11] and in accordance with [52-53], plasma BLs with a maximum diameter of up to 94 cm were created in the laboratory in crossed alternating electric and magnetic fields. Such BLs, as measurements showed, had spatially separated volumetric electric charges of different signs, with the motion of charges in circles, the electric currents of these charges and the corresponding magnetic field of dipole type. The distribution of radiation intensity, depending on the current radius from the center of the BL, is close in shape to the intensity distribution observed in natural BL. In this case, the maximum intensity is achieved near the BL shell, outside of which there is a rather quickly decaying luminous halo. Since the probes are positively charged upon contact with plasma BL, it is assumed that the BL shell contains relativistic protons, and the BL center contains electrons.

In electron-ion model, BL is actually a small piece of linear lightning, twisted into a ball with the typical size of 10 – 40 centimeters. Consequently, in both types of lightnings the currents and magnetic fields can be close in magnitude. Since the BL in this model has an overall positive charge, this may explain the fact that in [11] plasma BLs positively charged the electric charge probes. However, in the electron-ion model, the BL shell contains not protons, but electrons.

Confirmation of the electron-ion model of BL can be found in [54], where possible shape of magnetic field of BL was studied, which could ensure the motion of BL along sagging conductors, including power transmission lines. Taking into account properties of BL motion in a viscous air flow, it is concluded that electrostatic field of positively charged BL should be centrally symmetrical. The magnetic field of BL must have a symmetry axis, and the currents creating this magnetic field must be azimuthal. The equation of BL motion along a wire used in the calculations takes into account the wind pressure and interaction of the BL's magnetic field with electric current in the wire, which occurs due to charge leakage from BL surface into the wire. It is shown that such motion along the wire is possible only in the case when the currents generating the BL's magnetic field are concentrated in its surface layer. Indeed, the currents in volume of BL would interact with magnetic field in such a way that BL should increase its volume all the time, which is not observed. For configuration with the surface current, the expressions are found for components of vector potential and magnetic field of BL in spherical coordinates.

In [55-56] the hydrodynamic model of BL is considered in order to describe its motion through small hole. Outside the hole, equations of motion of an undeformed liquid ball in an ideal fluid flow of a point hydrodynamic source are used, which are written in spherical coordinates. These equations are modified into the Euler equation taking into account the continuity equation used to describe motion of BL in a hole, the diameter of which is by an order of magnitude smaller than the diameter of BL, which leads to transformation of the shape of BL into a jet. Solutions to these equations are found by numerical simulation under reasonable assumptions on the time of passage through the hole and on behavior of BL in air flow. One of the conclusions is that the shell of BL must have zero viscosity with respect to environment, and the matter of BL is an ideal incompressible fluid.

In [57], experiments were conducted with charged soap bubbles with a diameter of 3 cm, carrying an electric charge of the order of 10-15 nC. With such a charge, soap bubbles tend to be attracted by induced charges to a nearest grounded conductor. According to conclusions in [57], this casts doubt on possibility of charged BLs to move in horizontal direction and near grounded conductors if the BL's charge is large enough.

What can be said then regarding the possibility of horizontal motion of the BL, described in [26-27], which had the radius of 7 cm ? Based on the electron-ion model, the electric charge of such a BL is equal to $Q = 2.9 \times 10^{-7}$ C. According to calculations in [17], BL is subject to Archimedes' lifting force from the hot air inside BL, as well as to electric force acting between the BL's charge and induced charges on the ground. The equality of these forces is achieved when BL is at the height of about 10 cm above the ground, in which case BL can be attached to some object on the ground. If BL has the radius of 17 cm and corresponding charge $9.6 \times 10^{-6}$ C, then the equilibrium state will be achieved at the height of 90 cm.

Over time, the charge of BL can change due to interaction with environment or due to partial decay, leading to a change in equilibrium state. As the air inside BL cools down, its volume decreases, and with the loss of electrons from the outer shell charge $Q$ may also increase. Therefore, BL can rise above the ground and then move smoothly depending on the terrain configuration, the wind and electric fields from cumulonimbus clouds. Quite often, when transitioning from attached BL to free BL, it soars upwards and then goes along an inclined line towards the clouds. The latter can be explained by strong electric field acting on positive charge of BL from the negatively charged lower part of cloud. From equality of Archimedes' force and electric force, we can estimate the minimum electric field strength that can influence motion of BL:

$$E_m = \frac{\rho g V_b}{Q}. \tag{22}$$

For BL with radius of 7 cm and charge $Q = 2.9 \times 10^{-7}$ C, the electric field is equal to $E_m = 6.2 \times 10^4$ V/m. For comparison, field strength in stormy weather may well reach the value of $10^5$ V/m or more, based on potential difference between the clouds and the ground reaching $10^8$ V.

The comparison of charged soap bubbles and BL is not entirely correct, because these bubbles do not contain very hot air and are not subject to corresponding Archimedes force. In addition, we should consider interaction between the bubble's electric charge and massive conductor on the ground in more detail. In order for a bubble to be attracted to the conductor, the bubble's electric charge must separate the charges in conductor by its electric field. This requires some work to be done, which in this case is possible due to the gravity force acting on the bubble. The closer the bubble is to conductor, the more charges in it are separated and the stronger will be attraction to the conductor. In the absence of Archimedes' force, the balance of forces is such that the bubbles would be attracted to the conductors on the ground until they collide with them. In contrast, the motion of BL is controlled not only by the gravity force, but also by Archimedes' force, as well as by electrical forces from the charges on the ground and in the clouds and by the wind pressure. Therefore, the motion of BL can be more diverse, including various types of horizontal and vertical motion.

In [58], the electric charge and field potential near the BL's surface were determined in a very original way. For this purpose, the fact was used that BL, in accordance with the eyewitness evidence, was moving along a curved line approximately 1.5 m above the river surface along rafts made of logs with the initial angle of 30° to the wind direction. When describing one-dimensional motion of BL, the Navier-Stokes equation took into account the wind pressure, which depends on the height, and the electrostatic attraction, found using the image method, while the rafts made of logs were considered as dielectric protrusion above the water. As a result, with the radius of BL equal to 10 cm, the BL's charge was $Q = 5 \times 10^{-7}$ C, and the electric potential was 42 kV.

According to [59], the BL with a radius of 15 cm has a charge of the order of several microcoulombs and an electrical energy density of the order of one hundredth of a joule per cubic centimeter.

The BL parameters in [58-59] are in good agreement with our calculations of BL parameters in the electron-ion model in Section 3 and Table 1. In addition, according to calculations in [17], the electric energy of a BL with a radius of 17 cm is equal to 1.3 kJ. If we divide this energy by the volume of the BL, we obtain an electrical energy density of the order of 0.06 J/cm$^3$, which is consistent with the value of the electric energy density in [59].

## 7. Conclusion

The novelty of the electron-ion model compared to the proton-electron model in [10-11] lies in the fact that the outer shell of the BL contains not protons, but relativistic electrons, the currents of which generate strong magnetic field. This magnetic field captures the positive air ions inside the BL and causes them to rotate around the magnetic field lines. In turn, the internal electric charge of the BL keeps electrons as they rotate in the BL shell. This ensures long-term stability of the BL, which is generally positively charged. Charge separation in a BL occurs at the stage of its formation, with the key role played by the dipole magnetic field in the curved channel in the linear lightning branch. In case of strong current in linear lightning, the pinch effect becomes important.

The BL parameters found in Section 3, which depend on the BL radius, are confirmed by the observations described in Section 5, as well as by theoretical calculations and experiments presented in Section 6.

One of the advantages of the electron-ion model is that with its help we were able to explain not only the structure of ball lightning (BL), but also the structure of bead lightning, as well as occasionally observed structures of two coupled BLs. Unlike a single ball of spheroidal shape, in bead lightning there are several beads of almost spherical or cylindrical shape, interconnected by electric repulsive forces and magnetic forces of attraction. These forces, together with atmospheric pressure and hot plasma pressure, form the sizes of the emerging beads, in many cases exceeding diameter of primary linear lightning channel.

According to [1-2], beads often have a shape close to spherical. We can relate this to action of magnetic pressure, which arises due to electrons' rotation in the shell of each bead and generates a strong magnetic field inside the beads. According to observations, the greater are electron currents and magnetic field in the beads, the larger are sizes of the beads and energy stored in them, and the slower such beads cease to glow.

The beads occur in channel of normal linear lightning and therefore lose energy due to through currents in still conductive hot air of lightning channel. The heated air inside the bead is isolated by electron shell from the side surface, but can exit from the edges of the bead into lightning channel. As a rule, the lifetime of bead lightning significantly exceeds the glow time of primary linear lightning channel. It follows from observations that the beads can be seen for up to 2.5 seconds. As shown by calculations presented in Section 4, the electromagnetic forces holding the beads are so great that the beads can even withstand the pressure of a hurricane wind.

The electron-ion model can be useful for analyzing and explaining the results of experiments [60], in which luminous objects in the form of balls with dimensions of order of several millimeters were observed during an electric discharge in aqueous medium. During the discharge, water dissociates into hydrogen and oxygen ions that get inside these balls, which is followed by hydrogen burning. The fact that closely spaced balls merge with each other can be associated with the fact that magnetic force of attraction (13) at small distances between the balls is significantly greater than electric force of repulsion (12). When the balls collided with an obstacle, they exploded, which can be explained by stopping of charged particles moving rapidly in the shells of the balls, followed by ion recombination and corresponding rapid release of energy.

It should be noted that at the micro-level there could also be objects similar in their properties to BL and beaded lightning. Thus, in [61-62] we can find a review of properties of the so-called Exotic Vacuum Objects (EVO), which represent negatively charged spherical charge clusters ranging in size from 1 to 20 microns. The ratio of electrons to ions in EVO reaches $10^5$. EVOs were created in a low voltage / low power micro-arc discharge, they had velocities up to 1/10 of the speed of light and existed for up to $3\times10^{-11}$ s. An extraordinary property of EVO is their ability to create the similar chains as in beaded lightning. Just like BLs, EVOs can explode while producing X-ray emission. When encountering obstacles, EVOs make deep tunnels through them. To explain this property, in [63] EVOs are considered as objects in pulsed deposition systems and as shaped-charge munition. Among a number of questions that remain to be answered within the analogy between BL and EVO, the following can be mentioned: What forces cause EVO to form and maintain a spherical shape, despite the significant predominance of electrons over ions and repulsion of electrons from each other? How can EVOs be combined in chains considering their strong electrostatic repulsion from each other?

**Data availability**

The data that supports the findings of this study are available within the article.